\documentclass{article}
\usepackage{spconf,amsmath,graphicx}
\usepackage{hyperref, adjustbox, booktabs, amsfonts, aligned-overset, stmaryrd}
\usepackage{tabularx, multirow, multicol, makecell}
\usepackage{subcaption}

\usepackage{algorithm}
\usepackage{algpseudocode}
\usepackage{amssymb}
\usepackage{enumitem}
\usepackage[dvipsnames]{xcolor}
\newcommand{\Input}{\item[\textbf{Input:}]}
\newcommand{\Output}{\item[\textbf{Output:}]}

\title{SSAMBA: Self-Supervised Audio Representation Learning with Mamba State Space Model}
\name{Siavash Shams, Sukru Samet Dindar, Xilin Jiang, Nima Mesgarani}
\address{Department of Electrical Engineering, Columbia University, USA}

\begin{document}

\ninept
\maketitle
\begin{abstract}
Transformers have revolutionized deep learning across various tasks, including audio representation learning, due to their powerful modeling capabilities. However, they often suffer from quadratic complexity in both GPU memory usage and computational inference time, affecting their efficiency. Recently, state space models (SSMs) like Mamba have emerged as a promising alternative, offering a more efficient approach by avoiding these complexities. Given these advantages, we explore the potential of SSM-based models in audio tasks. In this paper, we introduce Self-Supervised Audio Mamba (SSAMBA), the first self-supervised, attention-free, and SSM-based model for audio representation learning. SSAMBA leverages the bidirectional Mamba to capture complex audio patterns effectively. We incorporate a self-supervised pretraining framework that optimizes both discriminative and generative objectives, enabling the model to learn robust audio representations from large-scale, unlabeled datasets. We evaluated SSAMBA on various tasks such as audio classification, keyword spotting, speaker identification, and emotion recognition. Our results demonstrate that SSAMBA outperforms the Self-Supervised Audio Spectrogram Transformer (SSAST) in most tasks. Notably, SSAMBA is approximately 92.7\% faster in batch inference speed and 95.4\% more memory-efficient than SSAST for the tiny model size with an input token size of 22k. These efficiency gains, combined with superior performance, underscore the effectiveness of SSAMBA’s architectural innovation, making it a compelling choice for a wide range of audio processing applications. Code at \textcolor{blue} {https://github.com/SiavashShams/ssamba.}  

\end{abstract} 

\section{Introduction}
\label{sec:intro}

Learning robust audio representations is critical for various tasks, including audio classification, speaker recognition, and emotion recognition \cite{mohamed_self-supervised_2022, Gong_Lai_Chung_Glass_2022}. Capturing both short-range and long-range dependencies is necessary for effective audio representation. While convolutional neural network models have shown limitations in capturing global dependencies, transformer models have excelled in image and language tasks due to their self-attention mechanisms \cite{vaswani2017attention, dosovitskiy_image_2021}. Building on these advancements, the Audio Spectrogram Transformer (AST) \cite{gong21b_interspeech} applied the self-attention mechanism to audio classification, achieving state-of-the-art performance in various audio classification benchmarks. AST training requires a large number of labeled audio clips, which can be difficult to find. To mitigate this, the Self-Supervised Audio Spectrogram Transformer (SSAST) \cite{Gong_Lai_Chung_Glass_2022} was introduced, employing an unsupervised pretraining framework. SSAST utilizes masked spectrogram patch modeling (MSPM) to pretrain the model on unlabeled audio data, significantly reducing the reliance on labeled data while maintaining competitive performance to AST.

\begin{figure}[!ht]
\centering
 \includegraphics[width=\columnwidth]{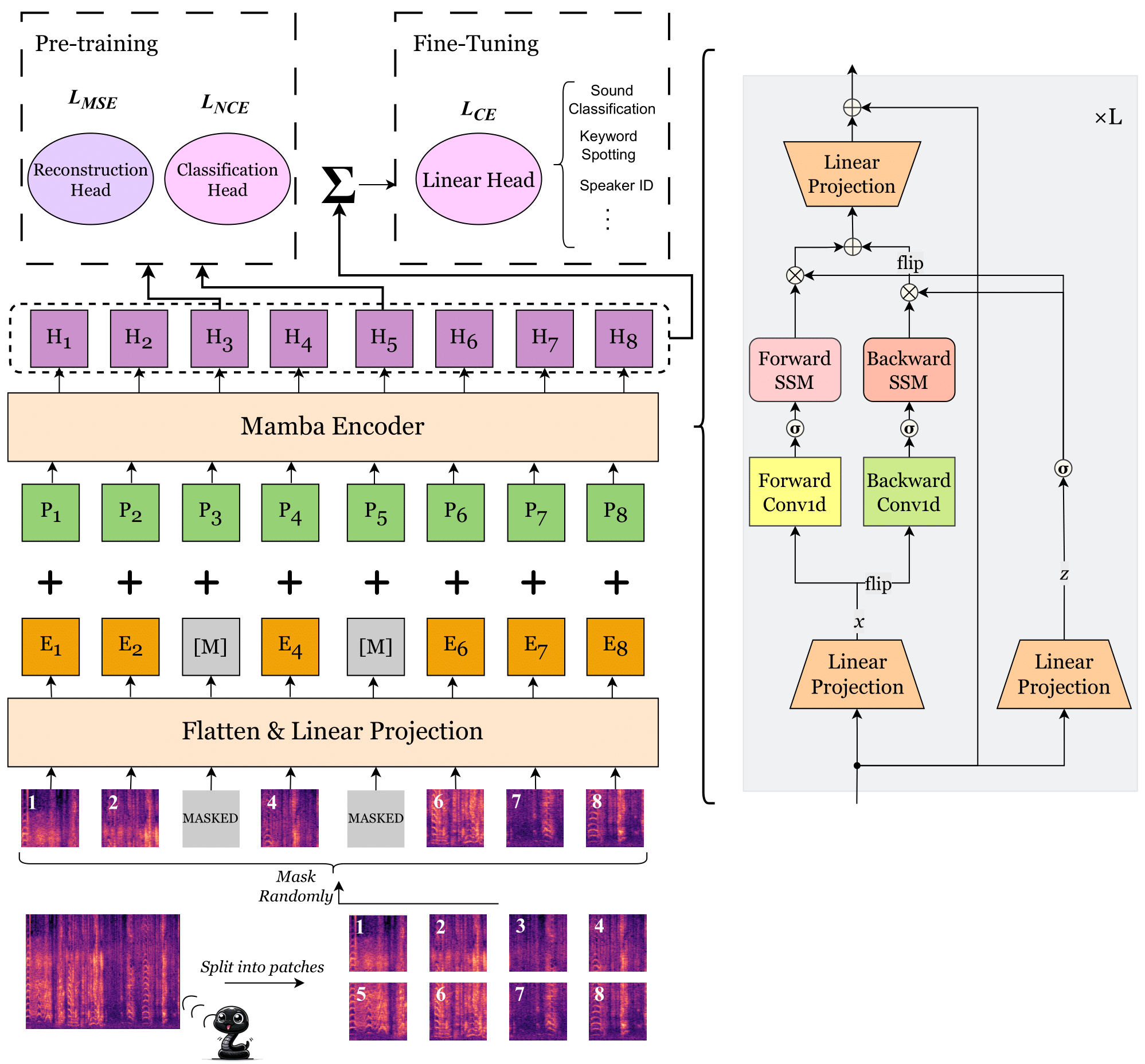}
 \caption{A top-down view of Self-Supervised Audio Mamba}
 \label{fig:audio mamba}
\end{figure}

Despite the high performance of SSAST, it still suffers from quadratic computation and memory usage due to its transformer architecture. As a more efficient alternative to transformers, state space models (SSMs)  \cite{kalman, albertgu1, albertgu2} have been explored in recent research. SSMs run with subquadratic complexity but maintain strong sequence modeling ability as transformers and can be trained in parallel. Notably, the newly proposed SSM model Mamba \cite{Gu2023Mamba} incorporates input-selective parameters within SSMs, improving sequence modeling ability while still enjoying linear complexity relative to sequence length. Its foundation paper demonstrates Mamba's high performance and efficiency in text, audio, and genomics modeling tasks. Subsequent studies have applied Mamba across a broader range of modalities and tasks. These models validate the Mamba's versatility and effectiveness in areas such as vision \cite{zhu_vision_2024, liu2024swin}, biomedical imaging \cite{ma2024umamba, xing2024segmamba}, video \cite{li2024videomamba}, and graphs \cite{wang2024graphmamba}.
Additionally, Mamba has been utilized in speech and audio applications. For instance, \cite{quan2024multichannel} employs Mamba for long-term multichannel speech enhancement. \cite{sui2024tramba} combines a hybrid transformer and Mamba model for acoustic and bone conduction speech enhancement. \cite{jiang_dual-path_2024} investigates speech separation using Mamba. These studies demonstrate that Mamba models can achieve performance comparable to transformer models.

Given the performance and efficiency demonstrated by these applications, Mamba has the potential to learn a general audio representation for multiple downstream tasks. Building on this potential, we propose Self-Supervised Audio Mamba (SSAMBA). In our approach, audio spectrograms are first split into patches and then transformed into an embedding sequence. These patches are subsequently fed into a bidirectional Mamba encoder, which captures the global audio context using selective state spaces. SSAMBA is trained with a self-supervised objective on masked spectrogram patches from a large unlabeled dataset. Once pretraining is complete, SSAMBA can be fine-tuned for specific downstream tasks using a small labeled dataset. Our experiments demonstrate that SSAMBA achieves superior or comparable performance to SSAST while significantly reducing inference costs.

The main contributions of this study are as follows:

\begin{itemize}
    \item We propose SSAMBA, the first self-supervised, attention-free, and SSM-based audio representation learning model. SSAMBA incorporates the bidirectional Mamba to encode and process audio, and it is pretrained without any labeled data.
    \item We implement and train SSAMBA in three sizes: Tiny, Small, and Base. All of them achieve similar or higher performance than the transformer model SSAST in downstream tasks: audio event classification, keyword spotting, speaker identification, emotion recognition, and dynamic audio scene labeling.
    \item We show the subquadratic-time computation and memory complexity of SSAMBA, which makes it a more efficient alternative to SSAST. For example, SSAMBA Tiny is approximately 92.7\% faster in inference speed and 95.4\% more memory-efficient than SSAST of the same size for the input length of 22k patches.
\end{itemize}

%%%% I updated to here (Siavash)

\section{Self-Supervised Audio Mamba}

In this section, we explore the mathematical foundations of the Mamba model, focusing on its state space model (SSM) framework and efficiency in capturing long-range dependencies. We describe the architecture of the Self-Supervised Audio Mamba (SSAMBA) model, which integrates bidirectional SSMs for robust audio context modeling. Finally, we explain the self-supervised learning framework adapted from SSAST \cite{Gong_Lai_Chung_Glass_2022}, utilizing masked spectrogram patch modeling (MSPM) to reduce reliance on labeled data \cite{Gong_Lai_Chung_Glass_2022}.

\subsection{Mathematical Foundations of the Mamba Model} \label{sec:ssm}

State space models (SSMs) are a powerful framework for sequence modeling, drawing inspiration from continuous systems that map a one-dimensional function or sequence \( x(t) \in \mathbb{R} \) to an output \( y(t) \in \mathbb{R} \) through a hidden state \( h(t) \in \mathbb{R}^N \). This is achieved using evolution parameters \( A \in \mathbb{R}^{N \times N} \) and projection parameters \( B \in \mathbb{R}^{N \times 1} \) and \( C \in \mathbb{R}^{1 \times N} \).

The continuous-time state space model is defined by the following differential equations:

\[ h'(t) = A h(t) + B x(t), \tag{1} \] 
\[ y(t) = C h(t).\tag{2} \]

To implement these models in digital systems, we need to discretize them. The discrete version of the SSM includes a timescale parameter \( \Delta \), which transforms the continuous parameters \( A \) and \( B \) to their discrete counterparts. The zero-order hold (ZOH) method is commonly used for this transformation, defined as follows:
\[ A_d = \exp(\Delta A), \tag{3}\]
\[ B_d = (\Delta A)^{-1} (\exp(\Delta A) - I) \cdot \Delta B. \tag{4}\]

After discretization, the state space model for a discrete-time signal with step size \( \Delta \) can be expressed as:
\[ h_t = A_d h_{t-1} + B_d x_t, \tag{5}\]
\[ y_t = C h_t. \tag{6}\]

To compute the output sequence \( y_t \) efficiently, we use a global convolution operation. The output \( y \) is obtained by convolving the input sequence \( x \) with a structured convolutional kernel \( K_{d} \), which is precomputed from the matrices \( A_{d} \), \( B_{d} \), and \( C \):
\[ K_{d} = (CB_{d}, CA_{d}B_{d}, \ldots, CA^{M-1}_{d}B_{d}), \tag{7} \]
\[ y = x \ast K_{d}, \tag{8} \]
where \( M \) is the length of the input sequence \( x \), and \( K \in \mathbb{R}^M \) is the structured convolutional kernel. The Mamba model enhances this framework by incorporating dynamic updates to the parameters \( \Delta_t \), \( A_t \), \( B_t \), and \( C_t \) based on the input \( x_t \) at each timestep \( t \). This makes the model input-selective and content-aware, allowing it to adjust to the specific characteristics of the input sequence dynamically. To efficiently handle these dynamic updates, Mamba employs a selective scan algorithm that recalculates the convolution dynamically, ensuring efficient and accurate sequence modeling.

\subsection{SSAMBA Architecture}

 Fig. \ref{fig:audio mamba} provides a comprehensive overview of the SSAMBA model, illustrating the following key components:

\subsubsection{Spectrogram Input Representation}
The input audio waveform is initially converted into a spectrogram, which represents the time-frequency domain of the audio data. This transformation is achieved by computing 128-dimensional log Mel filterbank features using a short-time Fourier transform (STFT) with a 25ms Hanning window applied every 10ms. The resulting spectrogram matrix \( S \) has dimensions \( 128 \times 100t \), where \( F = 128 \) is the number of frequency bins and \( T = 100t \) is the number of time frames for an audio length of \( t \) seconds. We then split this spectrogram into \( 16 \times 16 \) patches. For example, an audio input of 10 seconds, when divided into \( 16 \times 16 \) patches with a stride of 16, results in 500 patches.

\subsubsection{Flatten and Linear Projection}
Each spectrogram patch \( S_i \) is flattened into a 1D vector and projected into a higher-dimensional space using a linear projection layer. This results in embeddings \( E_i \), which have dimensions \( D \).

\subsubsection{Positional Encoding}
To capture the temporal order and spatial structure of the spectrogram patches, a learnable positional encoding \( P_i \) of the same dimension \( D \) is added to each patch embedding \( E_i \). This positional encoding ensures that the model retains the positional information of each patch within the spectrogram.

\subsubsection{Mamba Encoder}
The core component of SSAMBA is the Mamba encoder, which consists of bidirectional SSMs \cite{zhu_vision_2024}. The Mamba encoder processes the combined embeddings \( E_i + P_i \), capturing both forward and backward dependencies. This bidirectional approach processes information in both temporal directions, unlike unidirectional SSMs. The bidirectional SSM can be mathematically described as:

\begin{algorithm}
\caption{Bidirectional Mamba Block Processing}
\label{alg:bidirectional_mamba_block}
\begin{algorithmic}[1]
\Input Audio embedding sequence: $E_1, \ldots, E_M$ 
\Output Embeddings sequence: $H_1, \ldots, H_M$
\Statex

\For{$i = 1$ to $M$}
    \If{initial layer}
        \State $E'_i \leftarrow E_i + P_i$ \Comment{Add positional encoding to the initial layer input}
    \Else
        \State $E'_i \leftarrow H_i$ \Comment{For subsequent layers, use the output of the previous layer's corresponding patch}
    \EndIf
    
    \State $x \leftarrow \text{Linear}_x(E'_i)$
    \State $z \leftarrow \text{Linear}_z(E'_i)$

    \For{$o \in \{\text{forward}, \text{backward}\}$}
        \State $x'_o \leftarrow \text{SiLU}(\text{Conv1D}_o(x))$
        \State $B_o \leftarrow \text{Linear}_{B_o}(x'_o)$
        \State $C_o \leftarrow \text{Linear}_{C_o}(x'_o)$
        \State $\Delta_o \leftarrow \log(1 + \exp(\text{Linear}_{\Delta_o}(x'_o)))$
        \State $A_o \leftarrow \Delta_o \times \text{Parameter}_{A_o}$
        \State $B_o \leftarrow \Delta_o \times B_o$
        \State $y_o \leftarrow \text{SSM}(A_o, B_o, C_o)(x'_o)$
    \EndFor

    \State $y'_{\text{forward}} \leftarrow y_{\text{forward}} \odot \text{SiLU}(z)$
    \State $y'_{\text{backward}} \leftarrow y_{\text{backward}} \odot \text{SiLU}(z)$
    \State $H_i \leftarrow \text{Linear}_T(y'_{\text{forward}} + y'_{\text{backward}}) + E'_i$
\EndFor \\
\Return $H_1, \ldots, H_M$
\end{algorithmic}
\end{algorithm}

M represents the number of patches in the input sequence, and $z$ is an intermediate representation that modulates the forward and backward outputs of the SSM blocks.

\subsection{Self-Supervised Learning Framework}

The SSAMBA model employs a self-supervised pretraining framework designed to learn robust audio representations by jointly optimizing discriminative and generative objectives. This section details the key components and methodology of this framework.

\subsubsection{Masked Spectrogram Patches}
The spectrogram \( S \) is then split into a sequence of non-overlapping patches. Each patch \( S_i \) is of size \( F_p \times T_p \), where \( F_p \) and \( T_p \) are the dimensions of the patch in the frequency and time domains, respectively. During pretraining, a portion of these patches is randomly masked. The masked patches embeddings denoted by \( [M] \) are used as targets for the model to predict, forcing the model to learn the underlying structure of the audio data.

\subsubsection{Training Objective}

The training objective of SSAMBA integrates both discriminative and generative tasks to harness a comprehensive understanding of the audio spectrogram's structure. The overall training strategy involves two primary objectives:

\begin{itemize}[leftmargin=*]
    \item \textbf{Discriminative Objective:} This objective focuses on correctly identifying the masked patch. The discriminative task employs a classification head that outputs a vector for each masked patch, which is then compared against all other patch embeddings within the batch to compute the InfoNCE loss \cite{oord2018representation}:
    \[
    L_d = -\frac{1}{N} \sum_{i=1}^N \log \left(\frac{\exp(\langle c_i, x_i \rangle)}{\sum_{j=1}^N \exp(\langle c_i, x_j \rangle)}\right),
    \tag{9} \]
    where \( c_i \) is the output from the classification head for the \( i \)-th masked patch, \( x_i \) is the actual embedding of the \( i \)-th patch, and \( N \) is the total number of patches.

    \item \textbf{Generative Objective:} Alongside the discriminative task, the generative objective aims to reconstruct the original content of masked patches. A reconstruction head generates predictions for the masked embeddings, which are then evaluated using the Mean Squared Error (MSE) loss:
    \[
    L_g = \frac{1}{N} \sum_{i=1}^N \|\hat{x}_i - x_i\|^2,
   \tag{10} \]
    where \( \hat{x}_i \) is the predicted reconstruction of the masked patch and \( x_i \) is the true embedding of the patch.
\end{itemize}

The total loss \( L \) is a weighted sum of the discriminative and generative losses, with a balancing parameter \( \lambda \) controlling the relative importance of each:
\[
L = L_d + \lambda L_g. \tag{11} \label{eq:combined_loss}
\]

\section{Results} \label{results}

\begin{figure*}[htbp]
  \centering
  \begin{subfigure}[b]{0.48\textwidth}
    \centering
    \includegraphics[width=\textwidth]{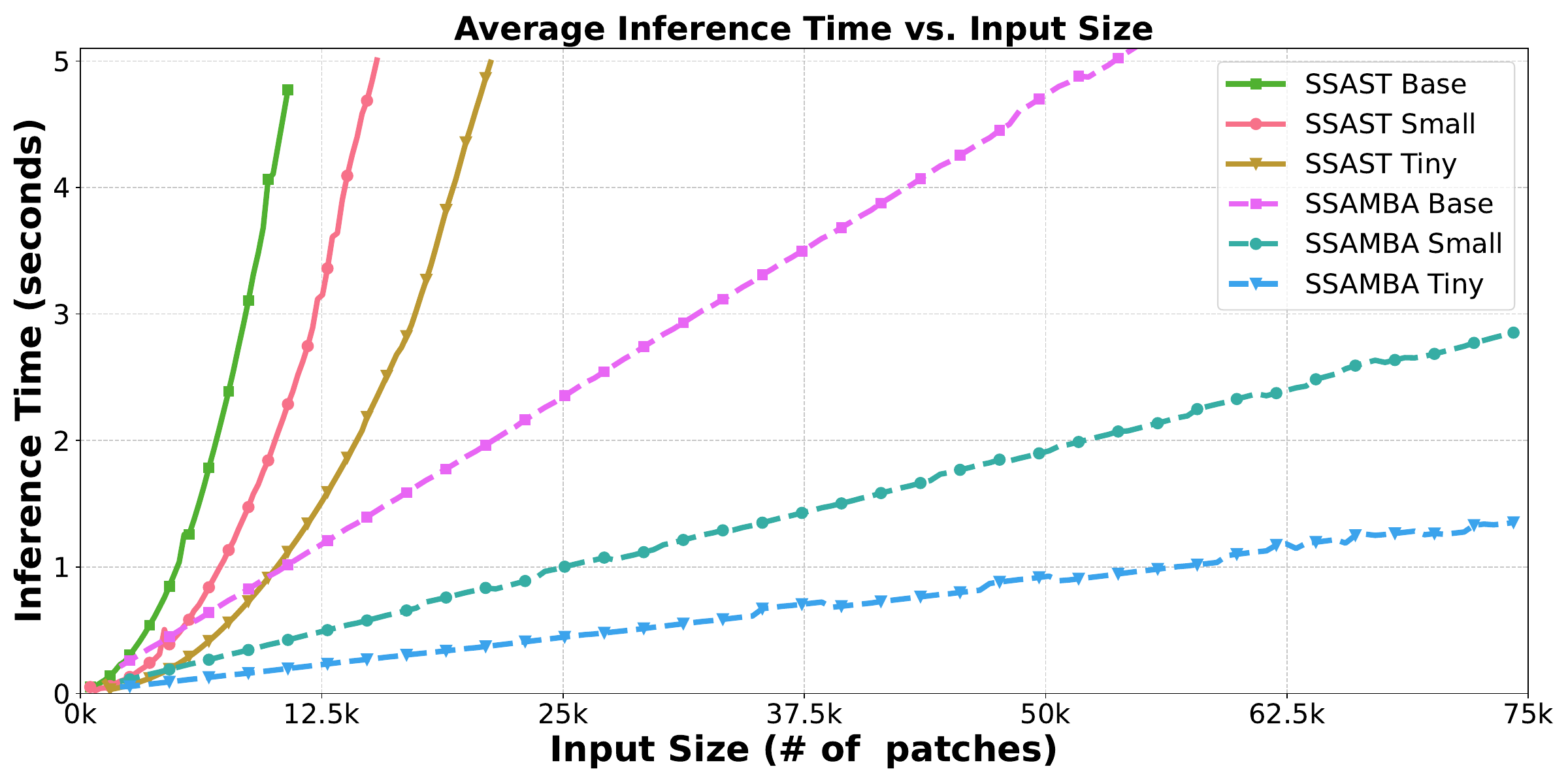} % Path to your inference time plot image
    \caption{Average Inference Time vs. Input Size}
    \label{fig:inference_time}
  \end{subfigure}
  \hfill
  \begin{subfigure}[b]{0.48\textwidth}
    \centering
    \includegraphics[width=\textwidth]{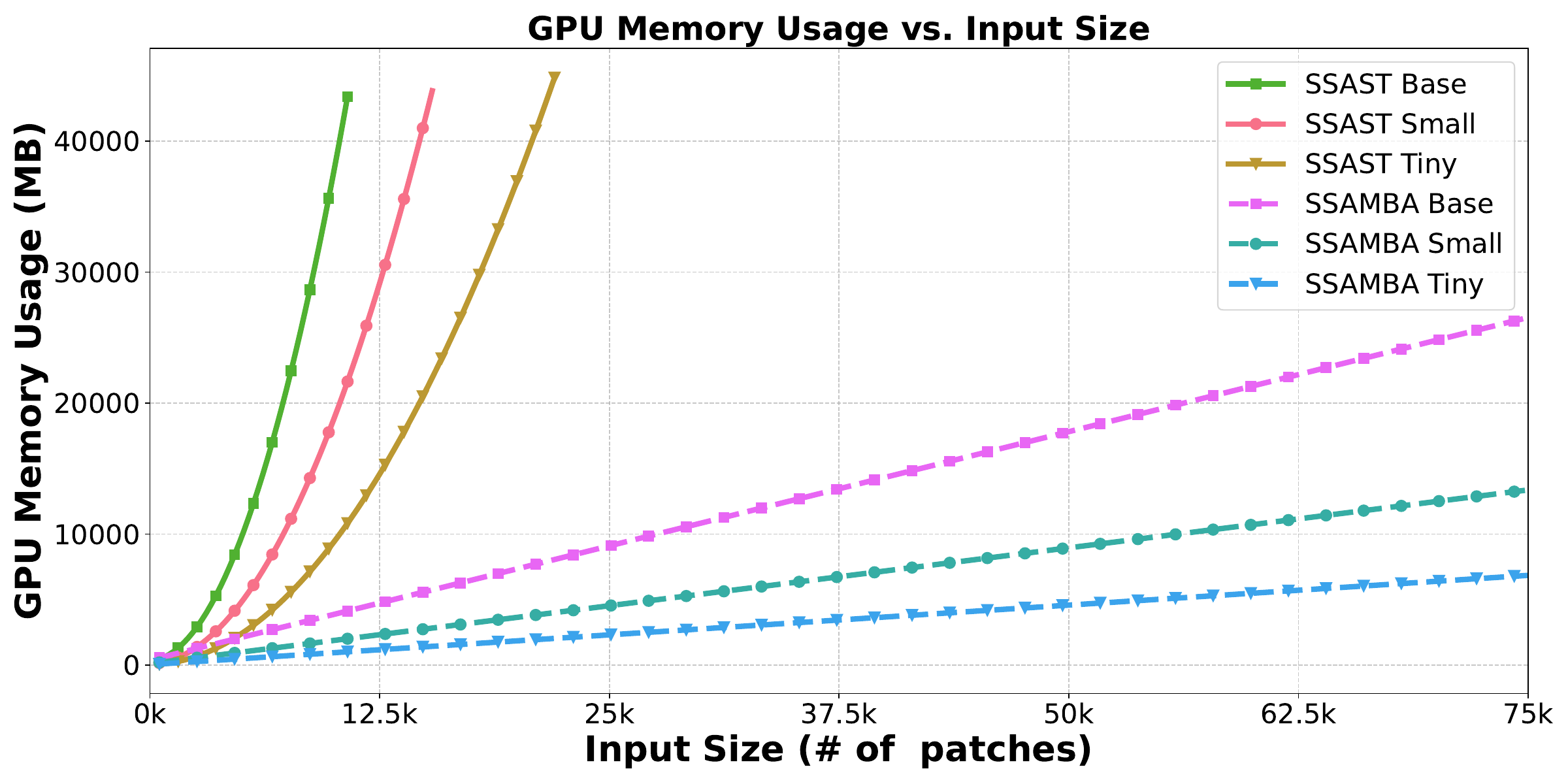} % Path to your GPU memory usage plot image
    \caption{GPU Memory Usage vs. Input Size}
    \label{fig:gpu_memory_usage}
  \end{subfigure}
  \caption{(a) Inference Time and (b) GPU Memory Usage for different model types and sizes}
  \label{fig:results}
\end{figure*}

\subsection{Pretraining }

For self-supervised pretraining of the SSAMBA model, we mixed and utilized audio samples from two datasets, focusing solely on the audio components and excluding any associated labels to foster a robust learning environment.
\begin{flushleft}
\textbf{Dataset Preparation and Integration:}
\end{flushleft}
\begin{itemize}[leftmargin=*]
    \item \textbf{AudioSet-2M:} We incorporated the entire unbalanced training set from AudioSet \cite{gemmeke2017audio}, which includes approximately 2 million 10-second audio clips from YouTube videos covering 527 distinct sound categories. These categories encompass a wide range of sounds from human and animal activities to natural and environmental noises. While AudioSet features a considerable presence of speech, it is often not the dominant component in the clips, which motivated the inclusion of an additional speech-focused dataset.

\item \textbf{LibriSpeech:} To better represent speech in the training data, we incorporated the 960-hour training set from LibriSpeech \cite{panayotov2015librispeech}. This dataset is composed of English audiobooks read by over 1,000 distinct speakers, offering a rich variety of speech patterns and accents. The inclusion of this dataset enhances the model’s ability to learn robust speech representations, which is particularly important given the wide range of audio tasks the SSAMBA model is designed to handle.

\end{itemize}
\begin{flushleft}
    \textbf{Data Handling and Training Configuration:}
\end{flushleft}
\begin{itemize}[leftmargin=*]
    \item To standardize the audio input, we processed all tracks from both AudioSet and LibriSpeech to ensure a uniform duration of 10 seconds per sample, either by cutting or padding the waveforms accordingly.
    \item We downsampled all audio files to 16kHz and converted stereo tracks to mono by averaging the channels, streamlining the audio input format for consistent model training.
    \item We employed the Adam optimizer \cite{adam} with a learning rate of \(1e-4\) and a batch size of 64. We used $\lambda = 10$ for the combined loss in Equation \eqref{eq:combined_loss}.
    \item Training was conducted on NVIDIA L40 GPUs. We pretrained the base model on 2 GPUs, and used 1 GPU for the small and tiny sizes.
    \item The pretraining was limited to 10 epochs, with an early stopping criterion based on the validation loss; specifically, training was halted if no significant improvement in the loss was observed during three consecutive evaluations. This approach helped in preventing overfitting and ensured efficient use of computational resources.
\end{itemize}

\begin{table}[!t]
\centering
\caption{Comparison of Model Specifications}
\label{tab:model_specs}
\begin{sc}
\begin{adjustbox}{max width=\columnwidth}
\begin{tabular}{lccc}
\toprule
\textbf{Model} & \textbf{Params} & \textbf{Depth} & \textbf{Embed Dim} \\
\midrule
SSAST-tiny & 6M & 12 & 192 \\
SSAST-small & 23M & 12 & 384 \\
SSAST-base & 89M & 12 & 768 \\
SSAMBA-tiny & 7M & 24 & 192 \\
SSAMBA-small & 26M & 24 & 384 \\
SSAMBA-base & 99M & 24 & 768 \\
\bottomrule
\end{tabular}
\end{adjustbox}
\end{sc}
\end{table}

\subsection{Performance Comparison of SSAMBA and SSAST Models}

\subsubsection{Downstream tasks and dataset}
We evaluated the SSAMBA and SSAST models on a comprehensive set of tasks to assess their effectiveness across various audio processing scenarios. These tasks encompass multi-label and single-label audio event classification, keyword spotting, speaker identification, emotion recognition, and dynamic audio scene labeling. Three different sizes of both models—Tiny, Small, and Base—were compared. The specifications of these models are presented in Table \ref{tab:model_specs}.

\begin{table*}[!t]
\centering
\caption{Performance Comparison of SSAST and SSAMBA Models with and without Pretraining ($\dagger$ indicates using a larger learning rate for the final classification head).}
\label{tab:combined_performance_comparison}
\begin{adjustbox}{max width=\textwidth}
\begin{tabular}{lcccccccc}
\toprule
\textbf{Model} & \textbf{Pretrained} & \textbf{AS (mAP)} & \textbf{KS1 (Acc.)} & \textbf{KS2 (Acc.)} & \textbf{ESC (Acc.)} & \textbf{SID (Acc.)} & \textbf{ER (Acc.)} & \textbf{DASL (mAP)} \\
\midrule
SSAST-tiny & Yes & 23.2$\dagger$ & \textbf{94.8} & \textbf{97.1} & 79.5 & 64.3 & 55.7 & \textbf{74.5} \\
SSAMBA-tiny & Yes & \textbf{23.3$\dagger$} & 94.0 & 94.0 & \textbf{80.1} & \textbf{66.1} & \textbf{56.3} & 71.9 \\
\midrule
SSAST-small & Yes & 25.0$\dagger$ & 95.4 & \textbf{97.8} & 85.4 & 67.0 & \textbf{58.7} & \textbf{77.1} \\
SSAMBA-small & Yes & \textbf{25.6$\dagger$} & \textbf{96.4} & 96.3 & \textbf{85.5} & \textbf{67.9} & 58.2 & 75.2 \\
\midrule
SSAST-base & Yes & 26.9 & 96.0 & \textbf{97.9} & 88.8 & 68.8 & 59.6 & 78.7 \\
SSAMBA-base & Yes & \textbf{28.3} & \textbf{96.9} & 97.4 & \textbf{89.3} & \textbf{70.1} & \textbf{61.5} & \textbf{80.8} \\
\midrule[0.5pt]
SSAMBA-tiny & No & 16.3$\dagger$ & 94.7 & 94.2 & 44.5 & 50.1 & 56.3 & 15.6 \\
SSAMBA-small & No & 18.2$\dagger$ & 95.7 & 95.7 & 61.9 & 51.5 & 57.1 & 12.9 \\
SSAMBA-base & No & 20.4 & 95.7 & 96.1 & 65.6 & 53.6 & 58.2 & 13.0 \\
\bottomrule
\end{tabular}
\end{adjustbox}
\end{table*}

The datasets used for these evaluations include:

\textbf{AudioSet-20K (AS)}: This dataset is used for multi-label audio event classification. The AudioSet-20K training set is a balanced subset of AudioSet-2M, containing 20,785 audio clips. 

\textbf{ESC-50 (ESC)}: This dataset is utilized for single-label audio event classification. It consists of 2,000 five-second environmental audio recordings categorized into 50 distinct classes \cite{piczak2015esc}.

\textbf{Speech Commands V1 (KS1)}: This dataset is employed for keyword spotting tasks. It comprises 64,727 one-second recordings of 10 common speech commands, along with additional classes for silence and unknown words to handle false positives \cite{warden2018speech}.

\textbf{Speech Commands V2 (KS2)}: Similar to KS1, this dataset is also used for keyword spotting but includes 105,829 one-second recordings of 35 common speech commands \cite{warden2018speech}.

\textbf{VoxCeleb 1 (SID)}: This dataset is used for speaker identification. It includes speech from 1,251 speakers, with the task being to classify each utterance by its speaker identity \cite{voxceleb}.

\textbf{IEMOCAP (ER)}: This dataset is employed for emotion recognition. It contains approximately 12 hours of emotional speech data, used to classify different emotional states \cite{iemocap}.

\textbf{Urban8K Sound (DASL)}: This task involves dynamic audio scene labeling using synthesized 1-minute audio sequences created by concatenating variable-length audio clips from the UrbanSound8K dataset, which includes 8,732 labeled sound excerpts of urban sounds from 10 classes (e.g., car horn, children playing, dog bark) \cite{salamon2014dataset}. 

During the fine-tuning phase, unlike the pretraining setup, no patches are masked. For standard tasks, the approach involves averaging all output tokens from the encoder before classification. However, for the Urban8K Sound (DASL) task, we use a modified method. Specifically, the output tokens corresponding to 1-second segments are averaged to align the model’s output with the temporal structure of urban sound events. Fine-tuning is then conducted based on the labels associated with each 1-second segment. To ensure the robustness of the model, we performed a 10-fold cross-validation, training on nine folds and testing on the remaining fold. The final results were averaged across all folds, providing a comprehensive evaluation of the model’s performance and consistency in handling longer audio sequences.

\subsubsection{Downstream Performance Comparison}

The results, summarized in Table \ref{tab:combined_performance_comparison}, illustrate that SSAMBA generally outperforms SSAST, particularly in the larger model configurations. SSAMBA's enhancements in architecture appear to provide superior handling of complex audio patterns, as evidenced by its consistently higher performance across the majority of tasks, especially in the Base model size. Notably, SSAMBA shows significant improvement in the AudioSet-20K and environmental sound classification tasks, suggesting robust feature extraction capabilities that scale well with model size. These improvements highlight SSAMBA's ability to effectively capture and leverage audio representations, resulting in better performance across diverse audio tasks. The table also emphasizes the significant advantages of pretraining, especially for complex tasks like DASL, where non-pretrained models struggle to generalize from training to validation data and fail to converge.

In this work, we primarily focused on comparing SSAMBA with SSAST, as both models share a similar pretraining methodology, making SSAST the most relevant baseline. This comparison allows us to directly contrast the fully transformer-based architecture of SSAST with the SSM-based architecture used in SSAMBA. 

To offer a broader evaluation, we also benchmarked SSAMBA against other leading self-supervised speech pretrainined models, including APC \cite{chung2019unsupervisedautoregressivemodelspeech}, Wav2Vec \cite{schneider2019wav2vec}, Wav2Vec 2.0 \cite{baevski2020wav2vec}, and HuBERT \cite{9585401}. The performance metrics for these models, sourced from the SSAST paper, are presented in Table \ref{tab:wav2vec2}.

In the SSAST paper, the authors also experimented with frame-based masking, a pretraining technique where the spectrogram is split into rectangular patches along the time axis. This approach allows the model to focus on temporal dynamics, which is particularly useful for speech-related tasks such as speaker identification (SID), emotion recognition (ER), and speech command recognition (SC). Frame-based masking has been shown to improve performance in these tasks by leveraging the sequential structure of speech \cite{Gong_Lai_Chung_Glass_2022}.

During the pertaining of SSAMBA, we employed patch-based masking as explained earlier. This method  is more effective for general audio representation learning. While not specifically optimized for speech tasks, SSAMBA still performs competitively against speech-specialized models. We anticipate that applying frame-based masking during the pretraining could further boost SSAMBA's performance in speech-specific tasks like SID, ER, and KS.

\subsubsection{Efficiency Comparison}
Efficiency is a critical factor for deploying deep learning models in real-world applications, where computational resources and inference times are often constrained. We compared the inference speed and GPU memory usage of SSAMBA and SSAST across different model sizes per varying input sizes, as depicted in Figure \ref{fig:results}. These comparisons were conducted with a batch size of 4 during inference.

The comparison shows that SSAMBA not only performs better in terms of accuracy but also offers significant improvements in efficiency. For instance, when comparing the Tiny models at an input size of 22k tokens, SSAMBA is approximately 92.7\% faster in inference speed and 95.4\% more memory-efficient than SSAST. These efficiency gains are crucial for real-time processing applications and deployment on resource-constrained devices.

\subsection{Ablations}

In this section, we explore the impact of varying the number of masked patches during training on the performance of different sizes of the SSAMBA model.  The primary goal of these ablations is to understand how different levels of input obfuscation during pretraining affect the model's ability to generalize and perform across various audio classification tasks. By systematically modifying the number of patches that are masked, we assess the robustness and flexibility of the model under varying degrees of information scarcity.

These experiments were conducted across three model sizes: Tiny, Small, and Base. Each model was tested with three different settings of masked patches—400, 300, and 250—to investigate how these variations influence performance metrics in audio event classification (AS-20K) \cite{gemmeke2017audio}, keyword spotting tasks (KS1 and KS2)\cite{warden2018speech}, Environmental Sound Classification (ESC)\cite{piczak2015esc}, and emotion recognition (ER) \cite{iemocap}. The results are shown in Table \ref{tab:masked_patches_performance}. All results in Table \ref{tab:combined_performance_comparison} are based on 400 masked patches.

For most tasks, the models trained with 400 masked patches perform better, particularly on general audio tasks such as AS and ESC. However, for speech-specific tasks like ER, we observe smaller performance gains as the number of masked patches decreases. These findings reflect the importance of optimizing the masking strategy based on the nature of the task.

\begin{table}[!t]
\centering
\caption{Performance Comparison of SSAMBA and Existing Speech Self-Supervised Pretraining Frameworks (* indicates frozen setting).}
\label{tab:wav2vec2}
\begin{adjustbox}{max width=\columnwidth}
\begin{tabular}{lccc}
\toprule
\textbf{Model} & \textbf{KS1 (Acc.)} & \textbf{SID (Acc.)} & \textbf{ER (Acc.)} \\
\midrule
APC \cite{chung2019unsupervisedautoregressivemodelspeech} & 94.0 & 60.4 & 59.3 \\
Wav2Vec \cite{schneider2019wav2vec} & 96.2 & 56.6 & 59.8 \\
Wav2Vec 2.0 \cite{baevski2020wav2vec}  * & 96.2 & 75.2 & 63.4 \\
HuBERT \cite{9585401} * & 96.3 & \textbf{81.4} & \textbf{64.9} \\
\midrule
SSAMBA-tiny & 94.0 & 66.1 & 56.3 \\
SSAMBA-small & 96.4 & 67.9 & 58.2 \\
SSAMBA-base & \textbf{96.9} & 70.1 & 61.5 \\
\bottomrule
\end{tabular}
\end{adjustbox}
\end{table}

\begin{table}[!t]
\centering
\caption{Impact of Masked Patches on Model Performance ($\dagger$ indicates using a
larger learning rate for the final classification head)}
\label{tab:masked_patches_performance}
\begin{sc}
\begin{adjustbox}{max width=\columnwidth}
\begin{tabular}{ccccccccc}
\toprule
\textbf{Size} & \textbf{Patches} & \textbf{AS} & \textbf{KS1} & \textbf{KS2} & \textbf{ESC} & \textbf{SID} & \textbf{ER} \\
& & (\textit{mAP}) & (\textit{Acc.}) & (\textit{Acc.}) & (\textit{Acc.}) & (\textit{Acc.}) & (\textit{Acc.})  \\
\midrule
\multirow{3}{*}{SSAMBA-Tiny}
& 400 & 23.3$\dagger$ & 95.7 & 94.0 & 80.1 & 66.1 & 56.3 \\
& 300 & 22.9$\dagger$ & 95.5 & 94.3 & 78.1 & 65.2 & 59.3\\
& 250 & 23.0$\dagger$ & 94.1 & 94.9 & 78.9 & 65.1 & 60.0 \\
\midrule
\multirow{3}{*}{SSAMBA-Small}
& 400 & 25.6$\dagger$ & 96.4 & 96.3 & 85.5 & 67.9 & 58.2\\
& 300 & 25.1$\dagger$ & 96.2 & 96.1 & 85.0 & 67.6 & 60.3\\
& 250 & 25.6$\dagger$ & 96.4 & 96.3 & 85.3 & 67.3 & 61.5 \\
\midrule
\multirow{3}{*}{SSAMBA-Base}
& 400 & 28.3 & 96.9 & 97.4 & 89.3 & 70.1 & 61.5\\
& 300 & 28.5 & 97.3 & 97.6 & 89.1 & 68.5 & 62.4\\
& 250 & 28.2 & 97.3 & 97.7 & 88.6 & 68.5 & 62.5 \\
\bottomrule
\end{tabular}
\end{adjustbox}
\end{sc}
\end{table}

In addition to varying the number of masked patches, we also explored other architectural choices, such as different normalization techniques and model directionality. Specifically, we compared RMSNorm with LayerNorm and experimented with fused add norm. However, these variations had little impact on the model’s performance. Similarly, we tested unidirectional Mamba encoder but found that it significantly underperformed compared to its bidirectional counterpart. As a result, we opted to proceed with the bidirectional configuration for our final evaluations.

\section{Conclusion}

In this paper, we introduced the Self-Supervised Audio Mamba (SSAMBA), a novel model for audio representation learning that leverages state space models (SSMs) with a bidirectional architecture. Unlike traditional transformer-based models with quadratic complexity, SSAMBA utilizes the Mamba architecture for greater efficiency and scalability. This is the first self-supervised, attention-free, SSM-based model applied to audio tasks. We validated SSAMBA through extensive experiments on various downstream tasks, including audio classification, keyword spotting, environmental sound classification, speaker identification, emotion recognition, and dynamic audio scene labeling. SSAMBA consistently outperformed the Self-Supervised Audio Spectrogram Transformer (SSAST), particularly in larger model configurations, and achieved significant efficiency improvements, with the Tiny model being 92.7\% faster and 95.4\% more memory-efficient than SSAST with an input size of 22k patches.

SSAMBA's robust performance is due to its architectural innovations and a self-supervised mixed dataset pretraining. SSAMBA’s efficiency on resource-constrained devices suggests potential for broad real-world applications, from mobile and edge devices to large-scale cloud systems.

\section{Acknowledgement}
    This work was funded by the National Institutes of Health (NIH- NIDCD) and a grant from Marie-Josee and Henry R. Kravis.


\bibliographystyle{IEEEbib}
\bibliography{strings,refs}

\end{document}